\documentclass[final,3p,times,twocolumn]{elsarticle}

\usepackage{hyperref}
\usepackage{graphics}
\usepackage{amssymb}
\usepackage{amsthm}
\usepackage{amsmath}
\usepackage{color}
\usepackage{natbib}
\usepackage{tikz}
\usepackage{pgfplots}
\usetikzlibrary{decorations.pathmorphing}
\usepgflibrary{decorations.text}
\usepgfplotslibrary{units}

\journal{Solid State Communications}

\begin{document}

\begin{frontmatter}

\title{Acoustic parametric pumping of spin waves}

\author[1]{Hedyeh Keshtgar}
\author[1]{Malek Zareyan}
\author[2,3]{Gerrit E. W. Bauer}

\address[1]{Institute for Advanced Studies in Basic Science, P. O. Box 45195-1159, 45195 Zanjan, Iran}
\address[2]{Institute for Materials Research and WPI-AIMR, Tohoku University, Sendai, 980-8577 Japan}
\address[3]{Kavli Institute of NanoScience, Delft University of Technology, 2628 CJ Delft, The Netherlands}

\begin{abstract}
Recent experiments demonstrated generation of spin currents by ultrasound. We can understand this acoustically induced spin pumping in terms of the coupling between magnetization and lattice waves. Here we study the parametric excitation of magnetization by longitudinal acoustic waves and calculate the acoustic threshold power. The induced magnetization dynamics can be detected by the spin pumping into an adjacent normal metal that displays the inverse spin Hall effect.
\end{abstract}


\end{frontmatter}

\section{Introduction}
Of particular recent interest in the field of spintronics are the two related effects of spin-transfer torque  \cite{BergerSTT, SlonczewskiSTT} and spin pumping \cite{TserkovnyakSP2002, TserkovnyakSP2005}. While the former refers to the torque exerted on the magnetization of a ferromagnet by an absorbed spin current with non collinear polarization, the latter deals with its inverse effect, in which spin angular momentum from the dynamics of the magnetization of a ferromagnet is transferred to the conduction electron spin of a paramagnetic metal contact \cite{BrataasSP&ST}.

The spin pumping requires magnetization motion that can be resonantly excited by microwaves with frequencies of the order of GHz. The effect has been already demonstrated in different magnetic structures utilizing metallic, semiconducting and insulating ferromagnets \cite{TserkovnyakSP2005,BrataasSP&ST}. In addition to the conventional way of a direct excitation, it is possible to parametrically excite spin waves with frequencies equal to an integer multiple of half the pump frequency. In the case of parallel pumping, in which magnetic field component of the exciting microwaves are parallel to the magnetization, the parametric excitation gives rise to a nonlinear absorption caused by the unstable growth of certain spin wave modes, as demonstrated by Schl\"{o}man  et al. \cite{Schlomann1} and Morgenthaler \cite{Morgenthaler60}. Here, the spin waves are excited at half of the frequency of the pump field. In the absence of damping any initial excitation will couple to the pump field and be amplified. Otherwise the spin wave amplitude grows non linearly when the power pumped into the system exceeds its power loss.

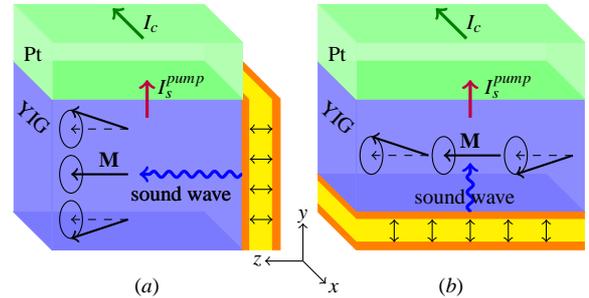
\begin{figure}[h]
\centering
\begin{tikzpicture}
\node[style={font=\footnotesize,thick}] at (1.75,-0.5){$(a)$};
\node[style={font=\footnotesize,thick}] at (5.75,-0.5){$(b)$};
\coordinate (Y1) at (0.5cm,0cm);\coordinate (Y11) at (4.5cm,0.5cm);
\coordinate (Y2) at (3cm,0cm);\coordinate (Y12) at (7.5cm,0.5cm);
\coordinate (Y3) at (0.5cm,2cm);\coordinate (Y13) at (4.5cm,2cm);
\coordinate (Y4) at (3cm,2cm);\coordinate (Y14) at (7.5cm,2cm);
\coordinate (Y5) at (0cm,0.5cm);\coordinate (Y15) at (4cm,1cm);
\coordinate (Y6) at (2.5cm,0.5cm);\coordinate (Y16) at (7cm,1cm);

\coordinate (P1) at (0.5cm,2.75cm);\coordinate (P11) at (4.5cm,2.75cm); 
\coordinate (P2) at (3cm,2.75cm);\coordinate (P12) at (7.5cm,2.75cm); 
\coordinate (P3) at (0cm,3.25cm);\coordinate (P13) at (4cm,3.25cm); 
\coordinate (P4) at (2.5cm,3.25cm);\coordinate (P14) at (7cm,3.25cm);
\coordinate (P5) at (2.5cm,2.5cm);\coordinate (P15) at (7cm,2.5cm); 
\coordinate (P6) at (0cm,2.5cm);\coordinate (P16) at (4cm,2.5cm);

\coordinate (A1) at (3.1cm,0cm);\coordinate (A11) at (4cm,0.9cm);
\coordinate (A2) at (3.4cm,0cm);\coordinate (A12) at (4cm,0.6cm);
\coordinate (A3) at (3.5cm,0cm);\coordinate (A13) at (4cm,0.5cm);
\coordinate (A4) at (2.6cm,2.5cm);\coordinate (A14) at (7.5cm,0.4cm);
\coordinate (A5) at (2.9cm,2.5cm);\coordinate (A15) at (7.5cm,0.1cm);
\coordinate (A6) at (3cm,2.5cm);\coordinate (A16) at (7.5cm,0cm);
\coordinate (A7) at (3.1cm,2cm);\coordinate (A17) at (4.5cm,0.4cm);
\coordinate (A8) at (3.4cm,2cm);\coordinate (A18) at (4.5cm,0.1cm);
\coordinate (A9) at (3.5cm,2cm);\coordinate (A19) at (4.5cm,0cm);

\filldraw[blue!70] (Y1) -- (Y2) -- (Y6) -- (Y5) -- cycle;
\filldraw[blue!50,opacity=0.9] (Y1) -- (Y2) -- (Y4) -- (Y3) -- cycle;
\filldraw[blue!50,opacity=0.9] (Y1) -- (Y3) -- (P6) -- (Y5) -- (Y1)-- cycle;
\filldraw[orange] (Y2) -- (Y4) -- (P5) -- (A4) -- (A7) -- (A1) -- cycle;
\filldraw[yellow] (A1) -- (A7) -- (A4) -- (A5) -- (A8) -- (A2) -- cycle;
\filldraw[orange] (A2) -- (A8) -- (A5) -- (A6) -- (A9) -- (A3) -- cycle;
\filldraw[green!30] (Y4) -- (P2) -- (P4) -- (P5) -- cycle;
\filldraw[green!30,opacity=0.9] (Y4) -- (P2) -- (P1) -- (Y3) -- cycle;
\filldraw[green!50] (Y3) -- (Y4) -- (P5) -- (P6) -- cycle;
\filldraw[green!40] (P1) -- (P2) -- (P4) -- (P3) -- cycle;
\filldraw[green!30,opacity=0.9] (Y3) -- (P1) -- (P3) -- (P6) -- cycle;

\filldraw[blue!70] (Y11) -- (Y12) -- (Y16) -- (Y15) -- cycle;
\filldraw[blue!50,opacity=0.9] (Y11) -- (Y12) -- (Y14) -- (Y13) -- cycle;
\filldraw[blue!50,opacity=0.9] (Y11) -- (Y13) -- (P16) -- (Y15) -- (Y11)-- cycle;
\filldraw[orange] (Y12) -- (Y11) -- (Y15) -- (A11) -- (A17) -- (A14) -- cycle;
\filldraw[yellow] (A11) -- (A17) -- (A14) -- (A15) -- (A18) -- (A12) -- cycle;
\filldraw[orange] (A12) -- (A18) -- (A15) -- (A16) -- (A19) -- (A13) -- cycle;
\filldraw[green!30] (Y14) -- (P12) -- (P14) -- (P15) -- cycle;
\filldraw[green!30,opacity=0.9] (Y14) -- (P12) -- (P11) -- (Y13) -- cycle;
\filldraw[green!50] (Y13) -- (Y14) -- (P15) -- (P16) -- cycle;
\filldraw[green!40] (P11) -- (P12) -- (P14) -- (P13) -- cycle;
\filldraw[green!30,opacity=0.9] (Y13) -- (P11) -- (P13) -- (P16) -- cycle;

\draw [<->] (3.1,0.4) --(3.4,0.4);\draw [<->] (5,0.1) --(5,0.4);
\draw [<->] (3.1,0.8) --(3.4,0.8);\draw [<->] (5.5,0.1) --(5.5,0.4);
\draw [<->] (3.1,1.2) --(3.4,1.2);\draw [<->] (6,0.1) --(6,0.4);
\draw [<->] (3.1,1.6) --(3.4,1.6);\draw [<->] (6.5,0.1) --(6.5,0.4);
\draw [<->] (7,0.1) --(7,0.4);
\draw[->,thick] (1.5,0.4) -- (0.75,0.15);\draw[->,thick] (7.35,1.25) -- (6.6,1);
\draw[->,dashed] (1.5,0.4) -- (0.75,0.4);\draw[->,dashed] (7.35,1.25) -- (6.6,1.25);
\draw (0.75,0.4) ellipse (0.15 and 0.25);\draw (6.6,1.25) ellipse (0.15 and 0.25);
\draw[->,thick] (1.5,1) -- (0.75,1);\draw[->,thick] (6.4,1.25) -- (5.65,1.25);
\node[style={font=\footnotesize}] at (1.25,1.2){$\textbf{M}$};
\node[style={font=\footnotesize}] at (6,1.45){$\textbf{M}$};
\draw (0.75,1) ellipse (0.15 and 0.25);\draw (5.65,1.25) ellipse (0.15 and 0.25);
\draw[->,thick] (1.5,1.6) -- (0.75,1.85);\draw[->,thick] (5.45,1.25) -- (4.7,1.5);
\draw[->,dashed] (1.5,1.6) -- (0.75,1.6);\draw[->,dashed] (5.45,1.25) -- (4.7,1.25);
\draw (0.75,1.6) ellipse (0.15 and 0.25);\draw (4.7,1.25) ellipse (0.15 and 0.25);
\draw [->,blue,very thick, decorate,decoration={snake,amplitude=.4mm,segment length=2mm,post length=1mm}] (6,0.5) -- (6,1.15);
\node[style={font=\footnotesize}] at (2.2,0.75){sound wave};
\draw [->,blue,very thick, decorate,decoration={snake,amplitude=.4mm,segment length=2mm,post length=1mm}] (3,1) -- (1.65,1);
\node[style={font=\footnotesize}] at (2.2,0.75){sound wave};
\node[style={font=\footnotesize}] at (5.93,0.7){sound wave};
\draw[->, purple,very thick] (1.75,1.75) --(1.75,2.25);
\node[style={font=\footnotesize}] at (2.2,2.15){$I^{pump}_s$};
\draw[->, purple,very thick] (6,1.75) --(6,2.25);
\node[style={font=\footnotesize}] at (6.5,2.15){$I^{pump}_s$};
\draw[->, green!50!black,very thick] (1.7,2.8) --(1.3,3.2);
\node[style={font=\footnotesize}] at (1.8,3){$I_{c}$};
\draw[->, green!50!black,very thick] (5.95,2.8) --(5.55,3.2);
\node[style={font=\footnotesize}] at (6.05,3){$I_{c}$};

\node[style={font=\footnotesize}] at (0.25,2.6){Pt};
\node[style={font=\footnotesize}] at (4.25,2.6){Pt};
\fill [decorate,decoration={text along path,text={|\footnotesize|YIG}}]
[draw] (0,1.85) -- (0.5,1.4);
\fill [decorate,decoration={text along path,text={|\footnotesize|YIG}}]
[draw] (4,1.85) -- (4.5,1.4);

\draw[->](3.8,-0.15) -- (3.8,0.35);
\node[style={font=\footnotesize}] at (3.8,0.45){$y$};
\draw[->](3.8,-0.15) -- (3.3,-0.15);
\node[style={font=\footnotesize}] at (3.2,-0.15){$z$};
\draw[->](3.8,-0.15) -- (4.1,-0.45);
\node[style={font=\footnotesize}] at (4.2,-0.5){$x$};
\end{tikzpicture}
\caption{Parametric excitation of spin waves by longitudinal acoustic wave propagating (a) parallel (b) perpendicular \cite{UchidaNM, UchidaASP} to the equilibrium magnetization vector along the $z$-direction. $I_s^{pump}$ and $I_c$ denote the spin pumping induced spin current into the adjacent normal metal in the $y$-direction and the charge current generated by the inverse spin Hall effect in the $x$-direction, respectively.}
\label{fig1}
\end{figure}

When there are no spin waves available at half the pump frequency, higher order instabilities occur. The critical field essential for parametric excitation increases with the order of instabilities. Accordingly, higher order instabilities are not observable as long as half the pump frequency lies in the accessible range of spin wave frequencies.

The exponentially growing amplitude of parametrically excited spin waves in time levels off by nonlinear effects mainly by interaction between the excited spin waves \cite{zakharov1975}.

More recently a mechanical type of spin pumping in ferromagnetic$\mid$normal metal (F$\mid$N) structures has been demonstrated, in which the magnetization motion is driven by injection of an acoustic wave into the F \cite{UchidaNM, UchidaASP, WeilerASP}. This acoustic spin pumping originates from  phonon-magnon coupling, i.e. the interaction between magnetization direction and elastic displacement. The linearized equations of motion in the presence of magnetoelastic coupling have been derived by Kittel \cite{Kittelme}. In the regime of linear coupling, the longitudinal elastic waves do not couple to the magnetization, however. Comstock \cite{Comstock1} has derived nonlinear equations of motion by retaining higher order terms in the magnetoelastic energy that can lead to the parametric excitation of the magnetization by the longitudinal elastic (pressure) waves.

Here we present a study of magnetoelastic coupling in a film of an insulating ferromagnet into which a longitudinal acoustic wave is injected. We demonstrate the possibility of parametric excitation of certain spin waves when the injected elastic waves propagating parallel or perpendicular to the applied bias field. We calculate the instability threshold and the corresponding critical value of the acoustic energy density flow. We also study the spin pumping induced DC spin current generated by the magnetization dynamics into an adjacent normal metal that displays the inverse spin Hall effect.

\section{Theory}
We consider a film of a magnetic insulator in the presence of an elastic wave. The starting point is the Landau- Lifshitz equation of motion for the magnetization vector $\textbf{M}(\textbf{r},t)$
\begin{equation}
\label{1}
\frac{\partial\textbf{M}(\textbf{r},t)}{\partial t}=-\gamma\mu\textbf{M}(\textbf{r},t)\times\textbf{H}_{eff}(\textbf{r},t),
\end{equation}
where $\gamma=\vert\gamma\vert$ is the gyromagnetic ratio, $\mu=4\pi\times10^{-7}\:\mathrm{Wb/\left(Am\right)}$ is the permeability and $\textbf{H}_{eff}(\textbf{r},t)=\textbf{H}_{ext}+\textbf{h}_p(r,t)+\textbf{H}_{dip}(\textbf{r},t)+\textbf{h}_{ex}(\textbf{r},t)$, the effective magnetic field consisting of the externally applied static field $\textbf{H}_{ext}$, an effective AC magnetic field $\textbf{h}_p$ caused by longitudinal acoustic wave and the internal magnetic fields $\textbf{H}_{dip}$ and $\textbf{h}_{ex}$ which arise due to the dipole-dipole and exchange coupling, respectively. The interaction of the longitudinal elastic waves with the magnetization is governed by the magnetoelastic energy. For cubic crystals,
\begin{align}
\label{2}
E_{me}=&\frac{b_1}{M_s^2}\left[m^2_x\varepsilon_{xx}+m^2_y\varepsilon_{yy}
+m^2_z\varepsilon_{zz}\right]\nonumber\\
&+2\frac{b_2}{M_s^2}\left[m_xm_y\varepsilon_{xy}+m_xm_z\varepsilon_{xz}
+m_ym_z\varepsilon_{yz}\right],
\end{align}
where $b_1$ and $b_2$ are the magnetoelastic coupling coefficients, $M_s$ is the saturation magnetization, $m_i$ are the magnetization components and the $\varepsilon_{ij}$ are components of the strain tensor, $\varepsilon_{ij}=\frac{1}{2}\left(\frac{\partial R_i}{\partial x_j}+\frac{\partial R_j}{\partial x_i}\right)$, with $R_i$ being the components of the displacement vector field $\textbf{R}(\textbf{r},t)$. The effective magnetic field corresponding to the above energy reads
\begin{equation}
\label{3}
\mu\textbf{h}_p(\textbf{r},t)=-\nabla_{\textbf{M}}E_{magnetoelastic}.
\end{equation}

In order to calculate the threshold of parametric excitation, we assume that the amplitudes of the spin waves at the threshold are small such that the nonlinear terms in Eq. (\ref{1}) can be neglected.  Far beyond the threshold, this assumption is not valid anymore. We introduce $\alpha_k(t)=\frac{1}{M_s}\left[m_{kx}(t)+i m_{ky}(t)\right]$ and $\alpha^*_{-k}(t)=\frac{1}{M_s}\left[m_{kx}(t)-i m_{ky}(t)\right]$ as the complex amplitude of spin waves in wave number ($k$) space \cite{Suhl1957}. Then, when the equilibrium magnetization lies in the plane of the film along the $z$-direction and the injected acoustic wave is longitudinal, the linearized equations of motion read 
\begin{align}
\label{7}
\frac{d}{dt}\begin{pmatrix}
\alpha_k(t)\\\alpha^*_{-k}(t)
\end{pmatrix}=i \begin{pmatrix}
A_k+c(t)&B_k+d(t)\\-B^*_k-d^*(t)&-A_k-c^*(t)
\end{pmatrix}\begin{pmatrix}
\alpha_k(t)\\\alpha^*_{-k}(t)
\end{pmatrix},
\end{align}
where
$A_k=\omega_H+\gamma\mu Dk^2+\frac{1}{2}\omega_M\sin^2\theta_k$, $B_k=\frac{1}{2}\omega_M\sin^2\theta_ke^{i2\phi_k}$ and $c(t)$ and $d(t)$ are determined by the effective ac field due to the elastic wave. Here $\theta_k$ and $\phi_k$ are, respectively, the polar and the azimuthal angles specifying the direction of the propagation vector $\textbf{k}=k\hat{\textbf{k}}$ of the spin waves, $D$ is the exchange constant, $\omega_H=\gamma\mu H_{ext}$ and $\omega_M=\gamma\mu M_s$.

A linear transformation which diagonalize Eq. (\ref{7}) in the absence of an external ac field gives,
\begin{align}
\label{9}
\frac{d}{dt}\begin{pmatrix}
b_k(t)\\b^*_{-k}(t)
\end{pmatrix}=i \begin{pmatrix}
\omega_k+F_k(t)&G_k(t)\\-G^*_k(t)&-\left[\omega_k+F_k(t)\right]^*
\end{pmatrix}\begin{pmatrix}
b_k(t)\\b^*_{-k}(t)
\end{pmatrix},
\end{align}
where $\omega^2_k=A^2_k-B^2_k$ is the spin wave frequency, $b_k=\lambda_k\alpha_k+\mu_k\alpha_k^*$, $F_k(t)=\lambda_k^2c(t)+\vert\mu_k\vert^2c^*(t)-\lambda_k\mu_k^*d(t)-\lambda_k\mu_kd^*(t)$ 
and $G_k(t)=-\lambda_k\mu_k\left[c(t)+c^*(t)\right]+\lambda_k^2d(t)+\mu_k^2d^*(t)$. Here, $\lambda_k=\sqrt{\left(A_k+\omega_k\right)/\left(2\omega_k\right)},$ and 
$\mu_k=\sqrt{\left(A_k-\omega_k\right)/\left(2\omega_k\right)}e^{i2\phi_k},$ are the Holstein-Primakoff transformation parameters \cite{HP}.

\section{Longitudinal elastic wave parallel to magnetization}
Here we consider a harmonic wave form $R_z=R\sin\left(\omega_pt-K_pz\right)$, with $R$, $K_p$ and $\omega_p$ being the amplitude, wavenumber and frequency, respectively. In the long wavelength limit, we obtain a spatially homogeneous ac field of the form
\begin{equation}
\label{5}
\textbf{h}_p(t)=\frac{2b_1}{\mu M_s}K_pR\cos\left(\omega_pt\right)\hat{z}.
\end{equation}
Accordingly, $c(t)=\gamma\mu h_p(t)$ and $d(t)=0$.
\subsection{First order instability }
\label{s1}
It is easy to show that the diagonal term $F_k(t)$ does not cause an instability but only modulates the frequency. Therefore we can disregard it in calculating the threshold of parametric excitation of spin waves.

An instability threshold requires dissipation in the magnetic system. The damping term in the Landau- Lifshitz-Gilbert equation of motion is given by $\frac{\alpha}{M_s}\textbf{M}\times\dot{\textbf{M}}$ \cite{GilbertLLG} where $\alpha$ is the Gilbert damping coefficient. In the small amplitude limit we can simply add an imaginary part to the spin wave frequency $\omega_k\rightarrow\Omega_k+i\eta_k$, 
where $\eta_k=\alpha A_k$ denotes the spin wave relaxation rate that may depend on $\textbf{k}$. Hence, Eq. (\ref{9}) can be rewritten as
\begin{align}
\label{10}
\frac{d}{dt}\begin{pmatrix}
b_k(t)\\b^*_{-k}(t)
\end{pmatrix}=i \begin{pmatrix}
\Omega_k&G_k^{(1)}e^{i\omega_pt}\\-G^{(1)*}_ke^{-i\omega_pt}&-\Omega_k^*
\end{pmatrix}\begin{pmatrix}
b_k(t)\\b^*_{-k}(t)
\end{pmatrix},
\end{align}
where we consider only the part of $G_k(t)$ that varies as $e^{i\omega_pt}$ (rotating wave approximation) and
\begin{equation}
\label{11}
G_k^{(1)}=-\frac{\omega_M}{4\omega_k}\sin^2\theta_ke^{i2\phi_k}\gamma\mu h_p.
\end{equation}
The parameter in the exponent of the trial solution $b_k(t)=b_ke^{i\frac{1}{2}\omega_p t+\kappa t}$ substituted in Eq. (\ref{10}),
\begin{equation}
\label{12}
\kappa=-\eta_k\pm\sqrt{\vert G^{(1)}_k\vert^2-(\omega_k-\omega_p/2)^2},
\end{equation}
reveals the instability condition, which is determined by the competition between acoustic energy  coupled  into the $b_k$ and represented by $G_k$, and the relaxation rate of the energy out of the $b_k$ represented by $\eta_k$. When $\kappa$ is positive the coupling dominates; the spin wave is unstable and the exponential growth of $b_k$ in time starts. Hence the threshold condition is given by $\kappa=0$.
The threshold for the dominant instability at $\omega_k=\omega_p/2$ is reached when $\vert G^{(1)}_k\vert=\eta_k$.
Therefore, the threshold amplitude $h_c$ of the effective pumping field for the spin wave pairs with $\pm\textbf{k}$ is obtained as \cite{Schlomann1, Morgenthaler60},
\begin{equation}
\label{14}
h_c=\frac{\omega_p}{\omega_M}\min\frac{\Delta H_k}{\sin^2\theta_k},
\end{equation}
where $\Delta H_k=2\eta_k/\gamma\mu$ is the spin wave resonance linewidth. The minimum value in Eq. (\ref{14}) is reached for a certain direction of the wave vector $\textbf{k}$. If we assume that $\Delta H_k$ is independent of $k$ and $\theta_k$, the spin waves with $\theta_k$ closer to $\pi/2$, i.e. normal to the phonon wave vector, are the first ones to become excited. With a low enough static field when $\omega_p/2>\omega_B$ (the frequency of spin waves with $k=0$ and $\theta_k=\pi/2$), the threshold is reached for spin waves with $\theta_k=\pi/2$. In this condition, the threshold of the elastic wave amplitude obtained from Eq. (\ref{5}) is
\begin{equation}
\label{15}
R_{c}=\frac{c\Delta H_k}{4\pi\gamma b_1}.
\end{equation}
Where $c$ is the sound velocity of pressure waves.

For Yttrium Iron Garnet (YIG) at room temperature, the density $\rho=5170\: \mathrm{kg/m^3}$, $\alpha=6.7\times10^{-5}$, $\gamma/2\pi=2.8\times10^{10}\:\mathrm{HzT^{-1}}$, $M_s=1.4\times10^5\:\mathrm{A/m}$, $c=7.2\:\mathrm{Km/s}$ and $b_1=3.5\times10^5\:\mathrm{J/m^3}$. The threshold amplitude, $h_c$, for a YIG in an external static field of magnitude $8\times10^4\: \mathrm{A/m}$ and effective microwave field of frequency $\omega_p/2\pi=10\:\mathrm{GHz}$ is about $40\:\mathrm{A/m}$ where $\Delta H_k\simeq20\:\mathrm{A/m}$. The corresponding density of the elastic energy flow, $\Pi=\rho\omega_p^2R_c^2c$, is $8\times10^5\:\mathrm{W/m^2}$. The threshold is 100 times smaller when $\omega_p/2\pi=1\:\mathrm{GHz}$ in agreement with the experiment has done by Matthews and Morgenthaler \cite{MatthewsEXP}.

The finite size effects of the film geometry can be taken into account by boundary conditions. They introduce another source of dissipation into the system which can not simply be added to $\eta_k$. These issues will be treated elsewhere. 

\subsection{Higher order instability}
At high static fields or low pump frequencies at which $\omega_p/2$ lies below the bottom of the spin wave spectrum, $\omega_H$, the first order instability does not occur and higher order instabilities can be observed, where the frequency of the unstable spin waves equals an integer multiple half the pump frequency.
To study these higher order instabilities, we use the identities $A_k=A_{-k}$ and $B_k=B_{-k}$ to transform Eq. (\ref{7}) to Hill's equation,  
\begin{equation}
\label{16}
\ddot{\alpha}+J(t)\alpha=0
\end{equation}
where we introduced a reduced time scale by replacing $\omega_p t\rightarrow2t$ and suppressed the subscript $k$. Here 
\begin{align}
\label{17}
J(t)=\left(\frac{2}{\omega_p}\right)^2&\lbrace\omega_k^2+\gamma\mu^2h_p^2\cos^2(2t)\nonumber\\
&+2\gamma\mu A_kh_p\cos(2t)+i\gamma\mu\omega_ph_p\sin(2t)\rbrace,
\end{align}
is a periodic function of period $\pi$ that can be written as $J(t)=\sum_{n=-2}^{n=2}J_ne^{2int}$ with
$J_0=\left(\frac{2\omega_k}{\omega_p}\right)^2+2\epsilon^2$, 
$J_1=2\epsilon\left(2\frac{A_k}{\omega_p}+1\right)$,
$J_{-1}=2\epsilon\left(2\frac{A_k}{\omega_p}-1\right)$ and
$J_2=J_{-2}=\epsilon^2$
where $\epsilon=\frac{\gamma h_p}{\omega_p}$ is defined as the pump ratio.
The general solution of Eq. (\ref{16}) has the form of $\alpha(t)=e^{\mu t}f(t)$, where $\mu=\xi+i\lambda$ is a complex quantity and $f(t)=\sum_{-\infty}^{+\infty}f_re^{2irt}$ is a periodic function of period $\pi$.
Accordingly, recursive set of equations is obtained.
\begin{equation}
\label{19}
J_2f_{r-2}+J_1f_{r-1}+\left[\left(\mu+2ir\right)^2+J_0\right]f_r+J_{-1}f_{r+1}+J_{-2}f_{r+2}=0
\end{equation}
Dividing the set of equations by $J_0-(2r)^2$, the determinant of the coefficient matrix reads,

\begin{align}
\label{20}
\Delta\left(i\mu\right)=
\begin{vmatrix}
&\vdots&\vdots&\vdots\\
\cdots&\frac{\left(i\mu+2\right)^2-J_0}{(2)^2-J_0}&\frac{-J_{-1}}{(2)^2-J_0}&\frac{-J_{-2}}{(2)^2-J_0}&\cdots\\
\cdots&\frac{J_{1}}{J_0}&\frac{\left(i\mu\right)^2+J_0}{J_0}&\frac{J_{-1}}{J_0}&\cdots\\
\cdots&\frac{-J_{2}}{(2)^2-J_0}&\frac{-J_{1}}{(2)^2-J_0}&\frac{\left(i\mu-2\right)^2-J_0}{(2)^2-J_0}&\cdots\\
&\vdots&\vdots&\vdots\\
\end{vmatrix}.
\end{align}
$\mu$ is obtained from the condition $\Delta\left(i\mu\right)=0$, which can be written in the form \cite{MclachlanHillEq}
\begin{equation}
\label{21}
\sin^2\left(\frac{\pi}{2}i\mu\right) =\Delta\left(0\right)\sin^2\left(\frac{\pi}{2}J_0^{\frac{1}{2}}\right).
\end{equation}
For the non-trivial solutions with $\xi\neq0$, Eq. (\ref{21}) reduces to
\begin{equation}
\label{22}
-\sinh^2\left(\frac{\pi}{2}\xi\right)=\Delta\left(0\right)\sin^2\left(\frac{\pi}{2}J_0^{\frac{1}{2}}\right),
\end{equation}
for the instability of even order and to
\begin{equation}
\label{23}
\cosh^2\left(\frac{\pi}{2}\xi\right)=\Delta\left(0\right)\sin^2\left(\frac{\pi}{2}J_0^{\frac{1}{2}}\right).
\end{equation}
for the instability of odd order. Joseph et al. \cite{JosephHOI} calculated $\Delta\left(0\right)$ up to the order of $\epsilon^4$ and derived the first and second order instability thresholds assuming $\epsilon\ll1$. 

For the first order instability, $J_0^{\frac{1}{2}}=1+\delta$ where $\delta$ indicates the small deviation of the resonant frequency of spin waves from half the pump frequency. The correction due to $\delta\neq 0$ makes a small modification to the threshold value. Therefore, in the limit of $\delta\rightarrow 0$, the same result as the previous subsection is obtained.

For the second order instability, the frequency of unstable spin waves is close to $\omega_p$ and $J_0^{\frac{1}{2}}=2+\delta$. According to Eq. (\ref{22}), $\xi^2\simeq-\delta^2+a\delta+b$ with $a=\frac{1}{12}J_1J_{-1}$ and $b=\frac{1}{16}J_2J_{-2}+\frac{\left(J_{-1^2}J_2+J_1^2J_{-2}\right)}{64}+\frac{5}{64\times36}J_{-1}^2J_1^2$ where $\delta\ll 1$, $\sinh\left(\frac{\pi}{2}\xi\right)\simeq\frac{\pi}{2}\xi$ and $\Delta\left(0\right)$ is calculated up to the order of $\epsilon^4$. The instability condition is fulfilled when the maximized $\xi$ with respect to $\delta$ equals to the dissipation rate, $2\eta/\omega_p$. Hereupon, the second-order instability threshold is \cite{JosephHOI}
\begin{equation}
\label{25}
h_c=\frac{\omega_p}{\gamma}\left(\frac{\Delta H_k}{2\pi M_s\sin^2\theta_k}\right)^\frac{1}{2}\left(1+\left(\frac{2\pi M_s\sin^2\theta_k}{\omega_p}\right)\right)^{-\frac{1}{4}}.
\end{equation}
As long as $\omega_p/2$ is in the range of accessible frequencies of spin waves ($\omega_p/2>\omega_H$), the first order process occurs and higher order instabilities are not observable. In general, higher order instabilities occur when the lower order instabilities are forbidden. For the same parameters as in the previous subsection, in order to suppress the first order instability, $H_{ext}$ has to be increased up to $15\times10^{4}\:\mathrm{A/m}$ or the pump frequency has to be reduced to $5\:\mathrm{GHz}$. In the former case, the second-order instability threshold occurs for an effective field $h_c\approx0.57\times10^{4}\:\mathrm{A/m}$, corresponding to an acoustic energy flux density of $4\times10^9\:\mathrm{W/m^2}$. While in the latter, the second-order instability threshold is at $h_c \approx0.23\times10^{4}\:\mathrm{A/m}$, or an elastic energy density flow of $7\times10^8\:\mathrm{W/m^2}$. 
Assuming that $\omega_p$ is in the accessible range of spin wave frequencies, the threshold is reduced with the pump frequency.

Higher order instability thresholds can be calculated in the same manner by retaining the higher order terms of pump ratio in $\Delta\left(0\right)$. For the instability of order $n$, the unstable spin wave frequency is close to $n\omega_p/2$ and $\Delta\left(0\right)$ should be calculated at least up to the order of $\epsilon^{2n}$. Thus, the threshold is proportional to $\left(\frac{2\eta}{\omega_p}\right)^{\frac{1}{n}}\omega_p$. Obviously, higher effective fields are needed to reach higher-order instability thresholds. 

\section{Longitudinal elastic wave perpendicular to magnetization}
Eq. (\ref{11}) implies that when $\theta_k=0$, there is no instability. Hence, the parametric excitation does not occur when the spin waves propagates along the magnetization. When the injected longitudinal elastic wave is transverse to the static field, say along the $y$ direction, $c(t)=-d(t)$ in Eq. (\ref{7}) and the phonon wavelength is long compared to the film thickness, $c(t)=\gamma\mu\frac{b_1}{M_s}K_pR\cos\left(\omega_pt\right)$. The coupling term in Eq. (\ref{9}) then reads,
\begin{equation}
\label{36}
G_k=\gamma\mu\frac{b_1}{M_s}RK_p\cos\left(\omega_pt\right),
\end{equation}
independent of the spin wave wave vector. Then, the first order elastic threshold is,
\begin{equation}
\label{37}
R_{c}=\frac{cM_s\Delta H_k}{b_1\omega_p}.
\end{equation}
The elastic energy density threshold of $4\times10^4\:\mathrm{W/m^2}$ for the parameters introduced in subsection \ref{s1} is lower for this configuration  (Fig. \ref{fig1}(b)) as compared to the previous one  (Fig. \ref{fig1}(a)).
  
\section{Spin pumping above threshold}
The spin wave amplitudes above the threshold instability can not grow unlimited. It is not the damping, but the interaction between pairs of waves with $\pm\textbf{k}$ is the main nonlinear mechanism which limits the amplitudes \cite{zakharov1975}. Therefore, in order to study the parametrically excited spin waves above the first order threshold, the four magnon interaction represented by the term $2i\sum_{k\prime}T_{kk\prime}a_ka_{k\prime}a_{k\prime}^*
+i\sum_{k\prime}S_{kk\prime}a_{-k}^*a_{k\prime}a_{-k\prime}$ should be inserted into Eq. (\ref{10}) written in terms of the slowly varying function of time, $a_k(t)=b_k(t)e^{-i\frac{\omega_p}{2}}$.
\begin{align}
\label{26}
\left(\frac{d}{dt}+\eta_k-i\left(\tilde{\omega}_k-\frac{\omega_p}{2}\right)\right)a_k-iP_ka^*_{-k}&=0\nonumber\\
\left(\frac{d}{dt}+\eta_k+i\left(\tilde{\omega}_k-\frac{\omega_p}{2}\right)\right)a^*_{-k}+iP^*_ka_k&=0
\end{align}
Where $\tilde{\omega}_k=\omega_k+2\sum_{k\prime}T_{kk\prime}n_{k\prime}$ is the renormalized spin wave frequency and $P_k=G_k^{(1)}+\sum_{k\prime}S_{kk\prime}\sigma_k\prime$ is the effective pumping. The newly introduced variables, $n_k=\langle a_\textbf{k}a^*_\textbf{k}\rangle$ and $\sigma_k=\langle a_\textbf{k}a_{-\textbf{k}}\rangle$ are the normal and anomalous spin wave correlation, respectively. 
We focus on the interesting case, when the amplitudes are equal in each parametric spin waves pair, $\mid a_{\textbf{k}}\mid=\mid a_{-\textbf{k}}\mid$. Hence, $\sigma_k=n_ke^{i\psi_k}$ where $n_k$ describes the intensity of spin waves pair with $\pm\textbf{k}$ which equals to the number of corresponding magnons and $\psi_k$ is the sum of phases of spin waves with $\textbf{k}$ and with $-\textbf{k}$.

From Eq. (\ref{26}) we obtain,
\begin{align}
\label{27}
\frac{1}{2}\frac{dn_k}{dt}&=\left[-\eta_k+\operatorname{Im}\left(P_k^*e^{i\psi_k}\right)\right]\nonumber\\
\frac{1}{2}\frac{d\sigma_k}{dt}&=\left[\left(\tilde{\omega}_k-\frac{\omega_p}{2}\right)+\operatorname{Re}\left(P_k^*e^{i\psi_k}\right)\right].
\end{align}
In the steady state above threshold, the spin wave pair with $\tilde{\omega}_k=\frac{\omega_p}{2}$ has the lowest threshold. This corresponds to $\operatorname{Re}\left(P_k^*e^{i\psi_k}\right)=0$ and $\mid P_k\mid=\eta_k$. Therefore, the total pair amplitude equals the number of parametrically excited magnons \cite{zakharov1975},
\begin{equation}
\label{28}
N=\sum_kn_k=\frac{\sqrt{\mid G_1^{(1)}\mid^2-\eta_1^2}}{S_{1,1}},
\end{equation}
and the phase of each pair is $\psi_1=\arcsin\left(\frac{\eta_1}{G_1^{(1)}}\right)$.
Here subscript $1$ indicates the first excited magnons which in the case of parallel pumping with $\frac{\omega_p}{2}>\omega_B$ refers to the spin waves with $\theta_k=\frac{\pi}{2}$. For these magnons, $S_{1,1}=\frac{1}{4}\left(\frac{\omega_M}{\omega_p}\right)^2\left(\sqrt{\omega_p^2+\omega_M^2}-\omega_M\right)$ is the amplitude of the four magnon interaction.

The magnetization dynamics in the system can be detected by the spin pumping into an adjacent normal metal that displays the inverse spin Hall effect. The DC component of the spin current pumped into the normal layer is given by \cite{TserkovnyakSP2002, TserkovnyakSP2005},
\begin{equation}
\label{30}
\textbf{I}_s^{pump}=\left<\frac{\hbar}{4\pi}\frac{g_r^{\uparrow\downarrow}}{M_s^2}\left[\textbf{M}\times\dot{\textbf{M}}\right]_z\right>_t
\end{equation}
where $g_r^{\uparrow\downarrow}$ is the real part of spin mixing conductance at the interface. The pumped spin current,
\begin{equation}
\label{31}
\textbf{I}_s^{pump}=\frac{\hbar}{8\pi}g_r^{\uparrow\downarrow}\omega_pN\hat{y},
\end{equation}
can be detected in the normal layer where it is converted into a charge current by the inverse spin Hall effect \cite{SaitohISHE},
\begin{equation}
\label{32}
\textbf{I}_{c}=\frac{2e}{\hbar}\theta_{SH}\mathbf{I}_s^{pump}\times\boldsymbol{\sigma}.
\end{equation}

Here, $\theta_{SH}$ is the spin Hall angle and $\boldsymbol{\sigma}$ is the direction of spin current polarization which for the dc component is in the direction of the magnetization precession axis, $\hat{z}$. Therefore, the charge current is in the $x$-direction.
\begin{figure}[h]
\begin{tikzpicture}
\begin{axis}[small,use units,
    x unit=W/m^2,x unit prefix=10^5,
    y unit=kA/m^3, 
        xlabel={Energy density flow},
        ylabel=$\mathrm{I_{c}}$]
\addplot[thick,smooth,samples=1000,color=blue,domain=0:10] {1.8*sqrt(x-8)};
\end{axis}
\end{tikzpicture}
\caption{The DC inverse spin Hall charge current as a function of acoustic energy density flow for $\omega_p/2\pi=10\:\mathrm{GHz}$. The threshold energy density flow is $8\times10^5\:\mathrm{W/m^2}$.}
\label{fiq2}
\end{figure}
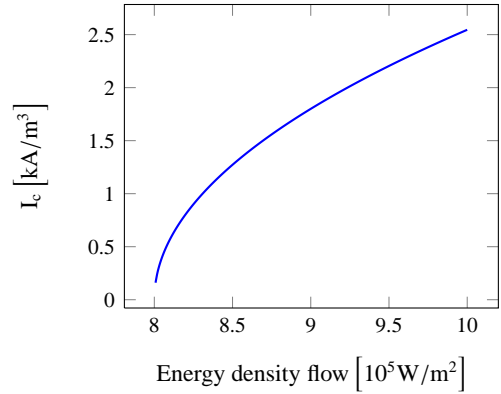
Fig. \ref{fiq2} demonstrates the dependence of charge current to the energy density flow when the normal layer is a Pt film with $\theta_{SH}=0.11$ and $g_r^{\uparrow\downarrow}=1\times10^{19}\:\mathrm{m^{-2}}$ \cite{WeilerSpnmixing}. Below the threshold value, no parametric excitation can take place. However, beyond that value, the amplitude of certain spin wave modes is amplified. The number of these parametrically induced magnons which have significant contribution in the pumped spin current is increased with the power pumped into the system.

\section{Conclusion}
In conclusion, we study the parametric excitation of spin waves by longitudinal acoustic waves propagating parallel and perpendicular to the bias field. The threshold of first and higher order instabilities are investigated and the corresponding critical ultrasound intensities have been found. Above the threshold, the DC spin pumped current generated by the magnetization dynamics into the adjacent normal metal and the induced detectable DC voltage is examined. The recent experiments by Uchida et al. \cite{UchidaASP} which demonstrate the generation of spin currents by sound waves of several MHz, i.e. far below the spin wave resonance frequencies, may be explained using the high order instability concept.

\section*{Acknowledgment}
We would like to thank Akash Kamra and Yaroslav Blanter for illuminating discussions. This work was supported by the FOM Foundation, Marie Curie ITN Spinicur, Reimei program of the Japan Atomic Energy Agency, EU RTN Spinicur, the ICC-IMR, DFG Priority Programme 1538 "Spin-Caloric Transport", and Grand-in-Aid for Scientific Research Nr. 25247056 and 252209. M. Z. thanks G. E. W. Bauer for the hospitality and support during his visit to the Institute for Materials Research of Tohoku University in Sendai.

\section*{References}
\bibliographystyle{elsarticle-num}
\bibliography{Mybib}

\end{document}